\documentclass[sigconf,nonacm]{acmart}

\AtBeginDocument{%
  \providecommand\BibTeX{{%
    \normalfont B\kern-0.5em{\scshape i\kern-0.25em b}\kern-0.8em\TeX}}}

\setcopyright{acmcopyright}
\copyrightyear{2018}
\acmYear{2018}
\acmDOI{10.1145/1122445.1122456}

\acmConference[]{}{}{}
\acmBooktitle{}
\acmPrice{}
\acmISBN{}

\usepackage[ruled,vlined]{algorithm2e}
\usepackage{amsmath}
\usepackage{float}

\begin{document}

\title{Correcting the User Feedback-Loop Bias for Recommendation Systems}

\author{Weishen Pan$^{1}$, Sen Cui$^{1}$, Hongyi Wen$^{2}$, Kun Chen$^{3}$, Changshui Zhang$^{1}$, Fei Wang$^{4}$}
\affiliation{%
	\institution{$^1$Institute for Artificial Intelligence, Tsinghua University (THUAI), State Key Lab of Intelligent Technologies and Systems, Beijing National Research Center for Information Science and Technology (BNRist), Department of Automation, Tsinghua University, P.R.China\\$^2$Cornell Tech, Cornell University, USA\\$^3$Department of Statistics, University of Connecticut, USA\\$^4$Department of Population Health Sciences, Weill Cornell Medicine, USA}
	\streetaddress{}
	\city{}
	\state{}
	\country{}
	\postcode{}
}
\email{{pws15,cuis19}@mails.tsinghua.edu.cn, hw557@cornell.edu}
\email{kun.chen@uconn.edu, zcs@mail.tsinghua.edu.cn, few2001@med.cornell.edu}

\renewcommand{\shortauthors}{}

\begin{abstract}
Selection bias is prevalent in the data for training and evaluating recommendation systems with explicit feedback. For example, users tend to rate items they like. However, when rating an item concerning a specific user, most of the recommendation algorithms tend to rely too much on his/her rating (feedback) history. This introduces implicit bias on the recommendation system, which is referred to as {\em user feedback-loop bias} in this paper. We propose a systematic and dynamic way to correct such bias and to obtain more diverse and objective recommendations by utilizing temporal rating information. Specifically, our method includes a deep-learning component to learn each user's dynamic rating history embedding for the estimation of the probability distribution of the items that the user rates sequentially. These estimated dynamic exposure probabilities are then used as propensity scores to train an inverse-propensity-scoring (IPS) rating predictor. We empirically validated the existence of such user feedback-loop bias in real world recommendation systems and compared the performance of our method with the baseline models that are either without de-biasing or with propensity scores estimated by other methods. The results show the superiority of our approach. 
\end{abstract}





\maketitle

\section{Introduction}
Recommendation has been a fundamental problem for many real-world applications ranging from e-commerce to healthcare. Many recommendation algorithms have been proposed in the last decade. Most of them tried to directly model explicit user feedback, such as the existing user-item ratings or user/item features. However, the implicit information hidden in recommendation systems can cause selection bias in the data and make the recommendation algorithms trained on such data not reliable. For example, on Netflix a specific user will only rate a movie if he/she has been exposed to that movie. Because of the large number of movies and the integration of recommendation systems, it is unlikely that each user has an equal opportunity to be exposed to every movie. If we construct a user-item rating matrix, this selection bias makes the missing values in the matrix not at random, which is referred to as the Missing-Not-At-Random (MNAR) problem \cite{marlin2009collaborative}. This makes the recommendation problem challenging.

To adjust for such selection bias, researchers have proposed different strategies in recent years to make the recommendations more reliable. For example, Schnabel {\em et al.} \cite{schnabel2016recommendations} proposed to estimate the quality of a recommendation system by propensity-weighting type of approaches \cite{imbens2015causal} and derived a matrix factorization approach to perform recommendation under the empirical risk minimization framework. Liang {\em et al.} \cite{liang2016modeling} proposed to explicitly model user exposure as a latent variable in the probabilistic matrix factorization framework \cite{mnih2008probabilistic} and developed an exposure matrix factorization algorithm for making recommendations. Bonner and Vasile \cite{bonner2018causal} proposed a domain adaptation approach to derive the causal embeddings of both users and items and estimate preference scores with them.

All of the research we mentioned above focused on static recommendation settings, i.e., no temporal rating information has been considered. In reality, many recommendation algorithms tend to rate the item for a specific user according to his/her rating history. It means the dynamics of the user ratings play an important role in the recommendation system and will introduce implicit bias, which we refer to such induced bias as {\em user feedback-loop bias}.

In this paper, we propose to systematically study the effect of user feedback-loop bias in recommendation systems and how to correct it in predicting item ratings. In particular, we design a novel probabilistic graphical model structure to estimate the dynamic/sequential item exposure probability distributions with respect to individual users. The estimated sequential exposure probabilities are then leveraged as propensity scores to adjust the user feedback-loop bias when we perform rating prediction for each specific item at each specific timestamp. We validate the existence of such user feedback-loop bias on the Movielens 20M and Goodreads dataset. We also demonstrate that better rating prediction performances and more diverse recommendation results can be achieved with the proposed dynamical de-biasing technique\footnote{We upload our source code on https://github.com/seq-ips/seq-ips.}.

\section{Related Work}
In this section, we briefly review the existing research that is related to this work.

\subsubsection{Causal Recommendation}
Different strategies have been proposed from the causal analysis perspective to make the recommendations more reliable. For example, researchers in \cite{liang2016modeling,wang2018modeling,wang2018collaborative} proposed to first estimate exposures for each user and then use them to de-bias click prediction, which has led to promising results. Other works include \cite{schnabel2016recommendations} and \cite{bonner2018causal} as mentioned above. These works focus on selection bias statically, while our paper studies selection bias as a dynamic process of how users rate the items over time.

\subsubsection{User Feedback-Loop in Recommendation Systems}

There are also recent studies on user feedback-loop and its influences on recommendation systems \cite{chaney2018algorithmic,nguyen2014exploring,schmit2018human,sharma2015estimating,shi2017long}. With simulations, Chaney {\em et al.} \cite{chaney2018algorithmic} demonstrated that the data generated from recommendation systems are confounded by the underlying recommendation algorithm, and with the continuous utilization of such system the algorithm tends to recommend homogeneous items to each user again and again at the end, which significantly decreased the users' satisfaction of the system. Schmit {\em et al.} \cite{schmit2018human} developed a model for analyzing user feedback-loop in recommendation systems and showed by simulation that that estimators which ignore the feedback-loop will deviate from the ideal recommendations. Other works are studying the effects of this bias by training collaborative filtering algorithms and sampling new data with the trained algorithm iteratively \cite{shi2017long}. And Sun {\em et al.} \cite{sun2019debiasing} develop an active learning method to debiasing the human-recommender system feedback loop. These works mainly illustrated via simulation the existence of user feedback-loop bias and its potential impacts in recommendation systems, however, such bias has not been validated in real-world data yet. In this paper, we empirically showed the existence of user feedback-loop bias in real recommendation systems and the correction of this bias would improve the quality of rate prediction, as evaluated by reduced prediction error and more diverse recommendation results.

\subsubsection{Sequential Recommendation}

Recently plenty of works have been proposed to model the sequential dependency of user behaviors. Some of them \cite{hidasi2015session,sachdeva2019sequential} model the item sequence with Recurrent Neural Network (RNN) and predict the next item a user will click, while some \cite{tang2018personalized,yuan2019simple} use Convolutional Neural Network (CNN) for sequence embedding. All these works are proposed to solve the implicit recommendation problem where the input is a sequence of items that a user has interacted before, and as such their models are not suitable to model user rating sequence, which is an essential component in our proposal.

\section{User Feedback-loop Bias and Problem Definition}

First, we use the framework in \cite{schnabel2016recommendations} to illustrate how to correct selection bias for rating prediction with a static exposure model. Then we introduce {\em user feedback-loop bias} and extend the framework to a sequential setting.

\subsection{De-biasing Learning of Rating Prediction}

Suppose there are $N$ users and $M$ items; let $Y \in \mathbb{R}^{N \times M}$ be the rating matrix, where $Y_{i,j}$ is user $i$'s rating on item $j$. Let $O \in \{0,1\}^{N \times M}$ indicate which items are rated by the users, that is, $O_{i,j}=1$ if $Y_{i,j}$ is observed. Let $P_{i,j} = {\rm P}(O_{i,j} = 1)$.

For the task of rating prediction, a rating estimator will be trained and produce the predicted ratings as $\hat{Y}$. Standard criterion to measure the performance of the estimator is as follows:

\begin{equation}
\nonumber
R(\hat{Y}) = \frac{1}{NM} \sum_{i=1}^{N}\sum_{j=1}^{M} \delta(Y_{i,j},\hat{Y}_{i,j}),
\end{equation}
where $\delta$ can have different forms such as Mean Squared Error (MSE): $\delta(Y_{i,j},\hat{Y}_{i,j}) = {(Y_{i,j} - \hat{Y}_{i,j})}^{2}$ and Mean Absolute Error (MAE): $\delta(Y_{i,j},\hat{Y}_{i,j}) = |Y_{i,j} - \hat{Y}_{i,j}|$ . Since $Y$ is not fully observed, in practice we can only measure $R(\hat{Y})$ on observed data:

\begin{equation}
\nonumber
\hat{R}(\hat{Y}) = \frac{1}{|(i,j):O_{i,j}=1|} \sum_{O_{i,j}=1} \delta(Y_{i,j},\hat{Y}_{i,j}).
\end{equation}

However, because of the existence of selection bias, $\hat{R}(\hat{Y})$ is no longer an unbiased estimation of $R(\hat{Y})$\cite{schnabel2016recommendations,steck2013evaluation}, which means that $\mathbb{E}_{O}[\hat{R}(\hat{Y})] \neq R(\hat{Y})$ ($\mathbb{E}_{O}$ is the expectation over the observed set $O$). To make the estimation unbiased, Schnabel used the Inverse-Propensity-Scoring (IPS) estimator \cite{imbens2015causal} to de-bias the learning process of recommendation systems by minimizing the following cost:

\begin{equation}
\nonumber
\hat{R}_{IPS}(\hat{Y}|P) = \frac{1}{NM} \sum_{O_{i,j}=1} \frac{\delta(Y_{i,j},\hat{Y}_{i,j})}{P_{i,j}}.
\end{equation}

The effectiveness of IPS estimator depends on a good estimation of $P_{i,j}$. Most of existing work build static exposure model to estimate $P_{i,j}$ \cite{liang2016modeling,schnabel2016recommendations} by considering different factors of selection bias. However, in real life, users rate items sequentially, and the system will change the recommendations according to users' rating history, making these models incapable of correcting such user feed-back loop bias.

\subsection{Selection Bias in a Sequential Manner}
\label{sec:selectionbiassequential}

As we stated above, recommendation systems rate the items and recommend customized items to the users according to the users' rating history in reality. Items have different probability to be rated since users typically rate items they like and rarely rate items they dislike. After each rating, the probability distribution of the items recommended by recommendation systems has changed because of the growing rating history. In other words, the dynamics of the user's rating history affect the work of recommendation systems and the sequential bias which the rating history leads to has to be considered.

\begin{figure}[htbp]
  \centering
  \includegraphics[width=3.0in]{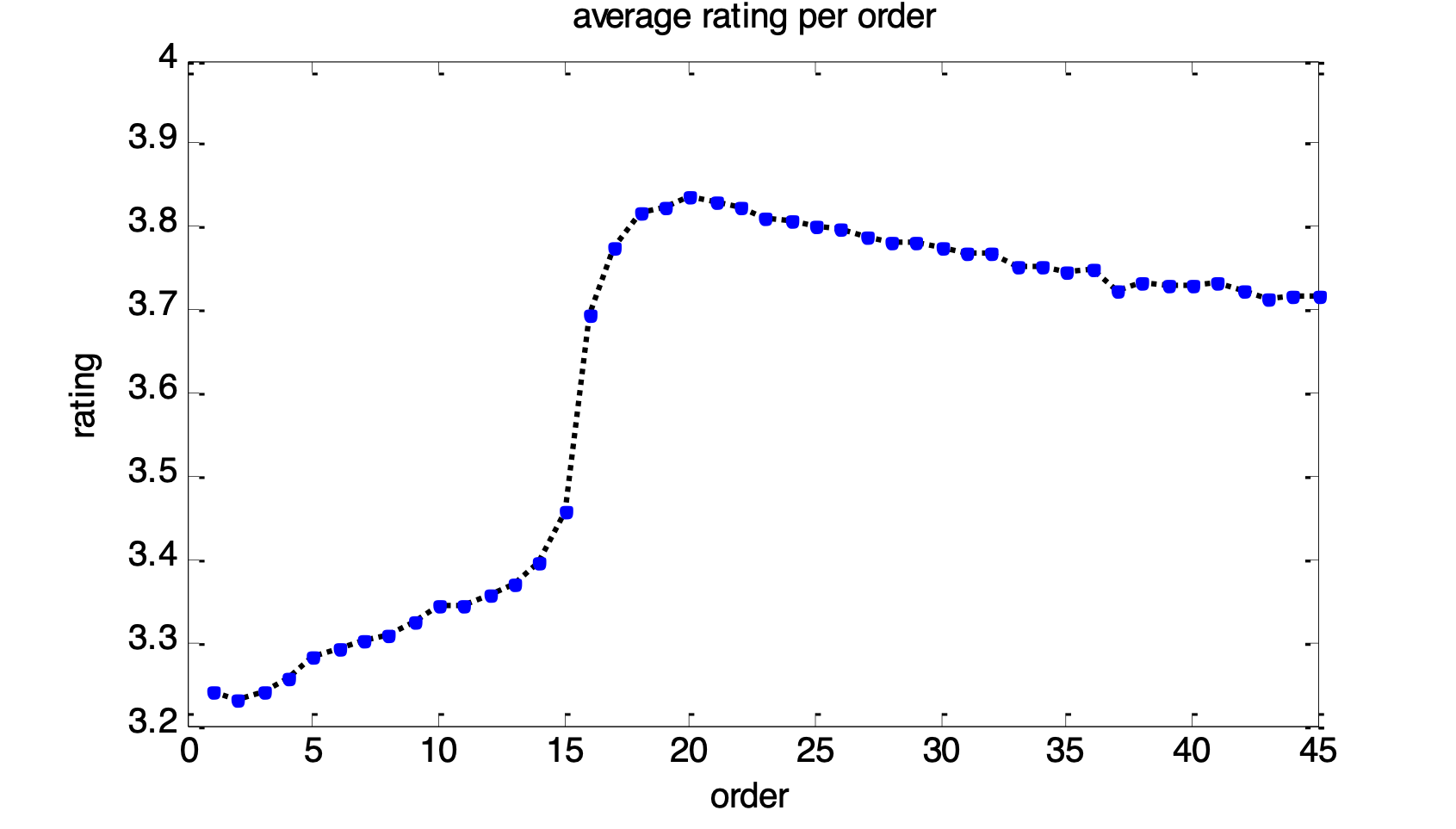}
  \caption{Average of each rating. The first 15 are the rates on items given by the recommendation system randomly respectively, and the rest are the rates on the movies recommended by the recommendation system according to the user's rating history}
  \label{fig:ratings}
\end{figure}

For further explanation, take an example in movie recommendation systems, users tend to rate movies with the styles they liked or popular recently, which are also the movies more likely to be exposed to the users. $Movielens-20M$ is a real-world data collecting ratings for movies by users (detailed description is in the section ~\ref{sec:movielens}). In which we select a part of the users' rating history. The mean ratings of the selected users are as Figure ~\ref{fig:ratings}.

In addition to the notations introduced above, the items rated by user $i$ is represented as a sequence $S_i = \{s^1_i,...,s^{|S_i|}_i\}$ where the index $k$ for $s^k_i$ denotes the item order. The observed ratings for user $i$ are {$Y_i|S_i = \{Y_{i,s^1_i},...,Y_{i,s^{|S_i|}_i}\}$}. $O^{k}_i \in \mathbb{R}^{M}$ is a one-hot exposure indication vector that $O^{k}_{i,j} = 1$ if $j = s^k_i$ and $O^{k}_{i,j} = 0$ otherwise. When user $i$ rates $s^k_i$, the corresponding rating history before such event is defined as $H_i^{1:k-1} = \{(s^1_i,Y_{i,s^1_i}),...,(s^{k-1}_i,Y_{i,s^{k-1}_i})\}$. $P^{k}_{i,j}$ is the probability that user $i$ rates item $j$ at order $k$.

Keep in mind that on Movielens website the recommendation system asks a new user to rate 15 items firstly, then the system will select customized movies based on the previous rating history. The first 15 rates can be regarded as unbiased rates respectively as they are not affected by any rates and all movies are likely to be recommended, while the other rates are affected by the previous rating history.

As illustrated in Figure ~\ref{fig:ratings}, the distribution of the first 15 rates is far different from the distribution of the rest rates. The first 15 rates are relatively low while the 16th rate increased a lot compared to the previous ratings. It means $P^{k>15}_{i,j}$ is extraordinary different from $P^{k<16}_{i,j}$. Since the recommendation system gets the information about the user's preference after 15 rates, it tends to recommend the movies that the user likes. After the first 15 rates, the rates are getting lower and lower. Because classic movies are the minority, we guess the system prefer recommending movies with higher ratings first. For the order $k>15$, the observed probability $P^{k}_{i,j}$ doesn't equals to $P^{k+1}_{i,j}$. Assuredly, the movies recommended after 15 movies are affected by dynamic {\em user feedback-loop bias} caused by the previous rating history.

Compared to existing research for adjusting the static selection bias, temporal user feedback-loop bias is dynamic as it is from the growing rating history.

\begin{equation}
\label{equ:obs_bias}
\begin{split}
\text{static model: }  \quad P^{k}_{i,j} = P_{i,j} \quad \text{for all k} \\
\text{dynamic model:  }  \quad P^{k_1}_{i,j} \not= P^{k_2}_{i,j} \quad \text{ if $k_1 \not= k_2$}
\end{split}
\end{equation}

Our main objective is to build a model to estimate $P^{k}_{i,j}$, which can be utilized to get an unbiased estimation of $R(\hat{Y})$. Here we assume the system conducts an algorithm to recommend items to a user based on rating history regularly. Specifically, with the sequential exposure probabilities, we propose the corresponding IPS estimator as (we use SIPS to indicate Sequential IPS):

\begin{equation}
\label{equ:sips_estimation}
\hat{R}_{SIPS}(\hat{Y}|P) = \frac{1}{NM} \sum_{i=1}^{N} \frac{1}{|S_i|} \sum_{k=1}^{|S_i|} \frac{\delta(Y_{i,s^k_i},\hat{Y}_{i,s^k_i})}{P^k_{i,s^k_i}}
\end{equation}
which is an unbiased estimation of $R(\hat{Y})$ because:
\begin{equation}
\begin{aligned}
\nonumber
&\mathbb{E}_{O}[\hat{R}_{SIPS}(\hat{Y}|P)] \\
&= \frac{1}{NM} \sum_{i=1}^{N} \frac{1}{|S_i|} \sum_{k=1}^{|S_i|}  \sum_{j=1}^{M} \mathbb{E}_{O^{k}_{i,j}}[\frac{\delta(Y_{i,j},\hat{Y}_{i,j})}{P^{k}_{i,j}} O^{k}_{i,j}] \\
&= \frac{1}{NM} \sum_{i=1}^{N} \frac{1}{|S_i|} \sum_{k=1}^{|S_i|}  \sum_{j=1}^{M} \delta(Y_{i,j},\hat{Y}_{i,j})\\
&= \frac{1}{NM} \sum_{i=1}^{N}\sum_{j=1}^{M} \delta(Y_{i,j},\hat{Y}_{i,j}).
\end{aligned}
\end{equation}

In fact, the selection bias comes from different sources. Though we focus on user feedback-loop bias, we also take into account other factors to get a more precise estimation of $P_{i,j}$. According to previous research \cite{baeza2016data}, other common bias factors include user activity, item popularity and self-selection bias (i.e., users are more likely to rate the items they like). User activity and item popularity will be included in our model. The user's self-selection bias is not considered \footnote{As for the user's self-selection bias, a completely random data set (which means the items are recommended to the users in a completely random way) is needed to estimate it well, which is very challenging. To the best of our knowledge, no public dataset with both random subset and sequential information is available.}. Based on the analysis above, we propose a novel dynamic exposure model to correct the sequential bias.

\section{Dynamic Exposure Model}

\begin{figure}[htbp]
  \centering
  \includegraphics[width=3.0in]{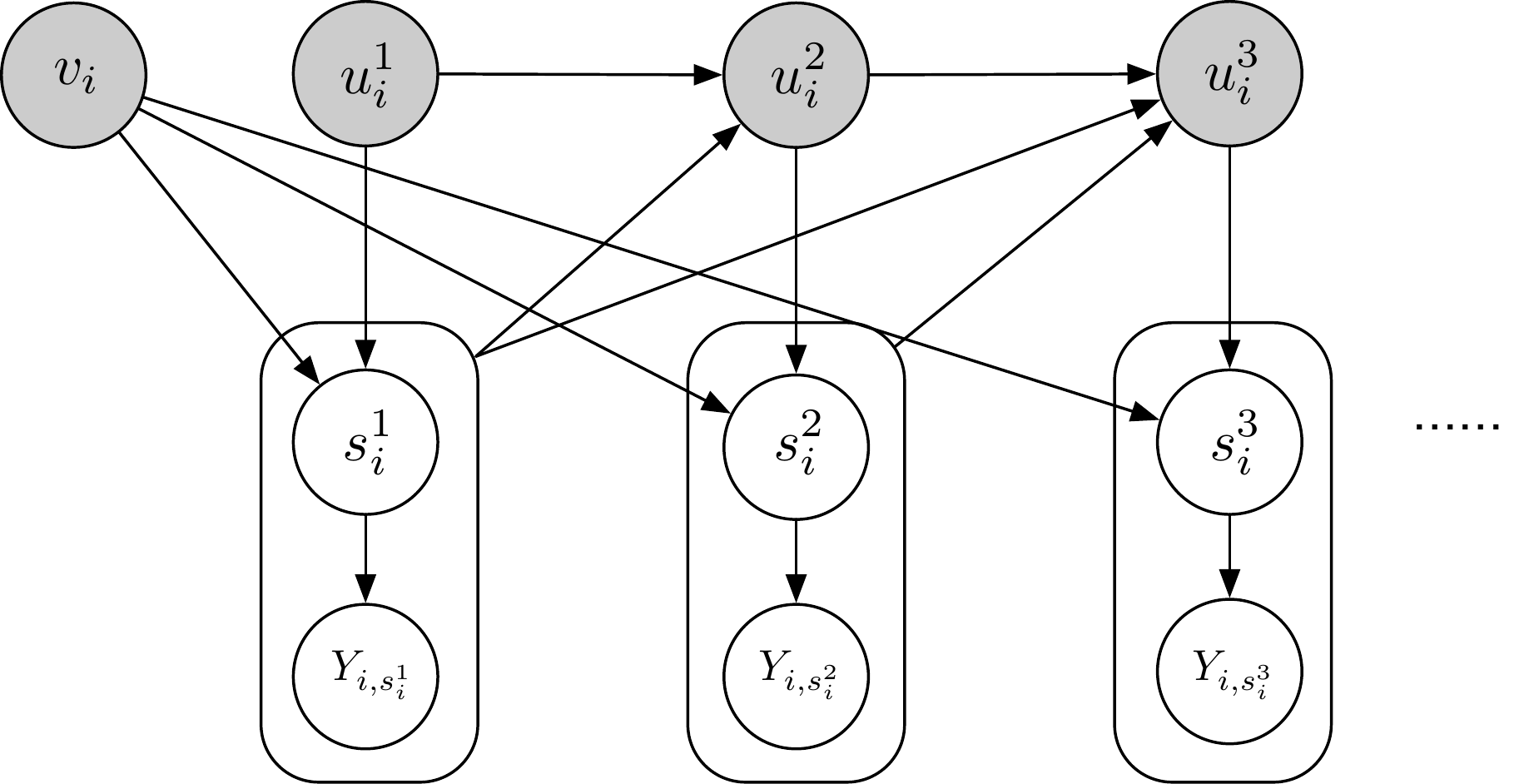}
  \caption{Dynamic exposure model. White nodes are variables which can be observed and collected, while the gray nodes are latent variables.}
  \label{fig:seq_expo_model}
\end{figure}

Figure. ~\ref{fig:seq_expo_model} provides an overview of our dynamical/sequential exposure model. We assume the exposure probability of each user is determined by two kinds of latent variables: the first one is the static $D_e$-dimensional latent factor $v_{i}$ sampled from a standard Gaussian prior. This variable is to represent the personal behavior/habit for a user to see and rate particular genres of items, which does not change along time:
\begin{equation}
\nonumber
p(v_{i}): v_{i} \sim N(0,I_{D_e}).
\end{equation}

The second one is a dynamic $D_e$-dimensional latent process indexed by $k$ as $u_{i}^{k}$. It represents how the system recommend items to a user based on rating history $H_{i}^{1:k-1}$:
\begin{equation}
\nonumber
p(u_{i}^{k}|H_{i}^{1:k-1}) :u_{i}^{k} \sim N(\mu_u(H_i^{1:k-1}),\Sigma_u(H_i^{1:k-1})),
\end{equation}
where $\mu_u$ and $\Sigma_u$ are functions of $H_i^{1:k-1}$. 

The $k^{th}$ item rated by user $i$ is sampled from a categorical distribution determined by $v_i$ and $u_{i}^{k}$:
\begin{equation}
\nonumber
p(s_i^k|v_{i},u_{i}^{k}): s^{k}_{i} \sim Cat(\pi(v_{i},u_{i}^{k})),
\end{equation}
where $\pi(v_{i},u_{i}^{k})$ is a $M$-dimensional probability vector.

Given the generative model, the posterior distribution of $v_{i},u_{i}^{1:|S_i|}$ is factorized as:
\begin{equation}
\begin{aligned}
\nonumber
& p(v_{i},u_{i}^{1:|S_i|}|S_i,Y_{i,S_i}) \\
& = p(v_{i}|S_i,Y_{i,S_i}) \prod_{k} p(u_{i}^{k}|H_{i}^{1:k-1},s_i^k,v_i).
\end{aligned}
\end{equation}

We directly mimic the structure of the posterior with the following factorization of the variational approximation:
\begin{equation}
\begin{aligned}
\nonumber
& q(v_{i},u_{i}^{1:|S_i|}|S_i,Y_{i,S_i}) \\
&= q(v_{i}|S_i,Y_{i,S_i}) \prod_{k} q(u_{i}^{k}|H_{i}^{1:k-1},s_i^k,v_i).
\end{aligned}
\end{equation}

We can get the variational lower bound of $\log p(S_i|Y_{i,S_i})$ as our loss function $\mathcal{L}$ to be optimized:
\begin{equation}
\label{equ:loss}
\begin{aligned}
&\log p(S_i|Y_{i,S_i}) \\
&\geq \mathbb{E}_{q(v_{i},\mathbf{u}_{i})} \log \frac{p(S_i,v_{i},u_{i}^{1:|S_i|}|Y_{i,S_i})}{q(u_{i}^{o},u_{i}^{h,1:|S_i|}|S_i,Y_{i,S_i})}\\
&= \mathbb{E}_{q(v_{i},\mathbf{u}_{i})} \log \frac{p(v_{i})\prod_{k}p(s_i^k|v_{i},u_{i}^{k})p(u_{i}^{k}|H_i^{1:k-1})}{q(v_{i},u_{i}^{1:|S_i|}|S_i,Y_{i,S_i})}\\
&= \mathbb{E}_{q(v_{i},\mathbf{u}_{i})} - KL(q(v_{i}|S_i,Y_{i,S_i}) || p(v_{i})) + \\
&\sum_{k}  (- KL(q(u_{i}^{k}|H_{i}^{1:k-1},s_i^k,v_i)|| p(u_{i}^{k}|H_{i}^{1:k-1})) + \\
&\log p(s_i^k|v_{i},u_{i}^{k})),
\end{aligned}
\end{equation}
where $q(v_{i},\mathbf{u}_{i})$ is short for $q(v_{i},u_{i}^{1:|S_i|}|S_i,Y_{i,S_i})$ and $KL$ represents KL-divergence.

\subsection{Network Structure} We now detail how we construct the generative model and variational approximation by neural networks.

The information of rating history $H_{i}^{1:k-1}$ is represented with a GRU (a kind of Recurrent Neural Network (RNN)), whose hidden state $h_i^{k}$ is updated by whose hidden state $h_i^{k}$ is updated with input of $s_i^{k-1}$ and $Y_{i,s_i^{k-1}}$:
\begin{equation}
\nonumber
\label{equ:gru}
h_i^{k} = GRU([W^{s}_{gru}s_i^{k-1},W^{y}_{gru}Y_{i,s_i^{k-1}}],h_{i}^{k-1}),
\end{equation}
where $W^{s}_{gru}$ and $W^{y}_{gru}$ are embedding matrices for one-hot item vector $s_i^{k-1}$ and its rate $Y_{i,s_i^{k-1}}$, $h_i^{k}$ is the $D_{gru}$-dimensional hidden state of GRU shared between generative and inference models.

\subsubsection{Generative Network}

For the generative process of $p(u_{i}^{k}|H_{i}^{1:k-1})$:
\begin{gather}
\nonumber
\label{equ:p_u}
\mu_u = W^{p}_{\mu_{u}}h_i^{k} + b^{p}_{\mu_{u}},\\
\nonumber
\Sigma_u = diag(\exp(W^{p}_{\Sigma_{u}}h_i^{k} + b^{p}_{\Sigma_{u}})).
\end{gather}
where $W^{p}_{\mu_{u}} \in \mathbb{R}^{D_{e} \times D_{gru}}$. The superscript $p$ means $W^{p}_{\mu_{u}}$ is the parameter of generative process and the subscript $\mu_{u}$ means $W^{p}_{\mu_{u}}$ is the mean parameter of variable $u_i^{k}$. $\mu_u$ and $\Sigma_u$ are short for $\mu_u(H_{i}^{1:k-1})$ and $\Sigma_u(H_{i}^{1:k-1})$. Similar rules apply to other parameters and functios defined below.

The exposure distribution $\pi(v_{i},u_{i}^{k})$ is calculated by:
\begin{equation}
\nonumber
\label{equ:expo_prob}
\pi(v_{i},u_{i}^{k}) = Softmax(W_{\pi} [v_{i},u_{i}^{k}] + b_{\pi}),
\end{equation}
where $W_{\pi} \in \mathbb{R}^{M \times 2D_{e}}$.

\subsubsection{Inference Network} 

For the posterior distribution $q(v_{i}|S_i,Y_{i,S_i})$, we use a normal distribution to approximate it, whose mean and variance are calculated as functions of $S_i$ and $Y_{i,S_i}$:
\begin{gather}
\nonumber
\label{equ:q_v}
\mu^{q}_{v} = \frac{1}{|S_i|} \sum_{k} W^{q}_{\mu_{v}}[s_i^{k-1},Y_{i,s_i^{k-1}}] + b^{q}_{\mu_{v}}\\
\nonumber
\Sigma^{q}_{v} = diag(\exp(\frac{1}{|S_i|} \sum_{k} W^{q}_{\Sigma_{v}}[s_i^{k-1},Y_{i,s_i^{k-1}}] + b^{q}_{\Sigma_{v}})).
\end{gather}

For the posterior distribution $q(u_{i}^{k}|H_{i}^{1:k-1},s_i^k,v_i)$, we also use a normal distribution to approximate it, whose mean and variance are calculated as functions of $H_{i}^{1:k-1},s_i^k,v_i$. Note that inference function of $u_i^{k}$ shares parameters with the corresponding generative model by using the same hidden state of GRU $h_i^{k}$ to calculate the posterior mean and variance as follows:
\begin{gather}
\label{equ:q_u}
\nonumber
\mu^{q}_{u} = W^{q}_{\mu_{u}}[h_i^{k},s_i^{k-1},v_i] + b^{q}_{\mu_{u}}\\
\nonumber
\Sigma^{q}_u = diag(\exp(W^{q}_{\Sigma_{u}}[h_i^{k},s_i^{k-1},v_i] + b^{q}_{\Sigma_{v}})).
\end{gather}

\section{Procedures of De-biasing Rate Prediction}
With the sequential exposure model, we use a two-step pipeline for de-biasing the rating prediction.

Firstly, we train a dynamic exposure model as in Algorithm \ref{alg:expo} where we use $\theta$ indicates all parameters of GRU, generative and inference networks and calculate the propensity $\hat{P}^k_{i,s_i^k}$ for each observed rating $Y_{i,s_i^k}$. We then use $\hat{P}^k_{i,s_i^k}$ as propensity scores to optimize for the evaluation function in equation~(\ref{equ:sips_estimation}) with MSE as loss function. We use Generalized Matrix Factorization (GMF) as rating prediction model, which is a generalized version of matrix factorization and achieve better performance on the validation set:

\begin{equation}
\nonumber
\hat{Y}_{i,j} = a_{out}(h^{T}(\alpha_i \odot \beta_j)),
\end{equation}
where ${\alpha_i} \in \mathbb{R}^{D_r}$ is the user-specific vector, ${\beta_j} \in \mathbb{R}^{D_r}$ is the item-specific vector, and $h \in \mathbb{R}^{D_r}$ and $a_{out}$ is the activation function.

\begin{algorithm}
\SetAlgoLined
\caption{Train the dynamic exposure model and estimate the exposure probability}\label{alg:expo}
\KwIn{$S_i$ and $Y_{i,S_i}$ for all users}
\KwOut{$\hat{P}_{i,s_i^k}$ for all users and their rated items}
Initialize $\theta$\\
\While{not converged}{
\ForAll{user i}{
Calculate {$h_i^{k}$}\\
Calculate $\mu_u(H_i^{1:k-1})$,$\Sigma_u(H_i^{1:k-1})$ for generative model\\
Calculate $\mu^{q}_{v}(S_i,Y_{i,S_i})$,$\Sigma^{q}_{v}(S_i,Y_{i,S_i})$, $\{\mu^{q}_{u}(H_{i}^{1:k-1},s_i^k,v_i),\Sigma^{q}_u(H_{i}^{1:k-1},s_i^k,v_i)\}$\\
Sample $v_i$ and $\{u_i^k\}$ from inference model\\
Estimate Monte-Carlo approx. to $\nabla_{\theta} \mathcal{L}$ and update $\theta$\\
}
}

\ForAll{user i}{
Calculate {$h_i^{k}$}\\
Calculate $\mu^{q}_{v}(S_i,Y_{i,S_i})$ and $\{\mu^{q}_{u}(H_{i}^{1:k-1},s_i^k,v_i)\}$ as $v_i$ and $\{u_i^k\}$\\
Compute $\hat{P}_{i,s_i^k}$ for each $s_i^k \in S_i$ and store
}
Return $\hat{P}_{i,s_i^k}$ for all users and their rated items
\end{algorithm}

\subsection{Implementation Details}

For each method, the grid search is applied to find the optimal settings of hyperparameters using the validation set. For both tasks, hyperparameters include all related dimensions $D_e$, $D_{gru}$ and $D_r$ from $\{16,32,64\}$, l2-normed regularization hyperparameter $\lambda_e$ for exposure model and $\lambda_r$ for rating model, learning rate $lr$ from $\{10^{-1},...,10^{-4}\}$, batch size for rating model $B$ from $\{64,128,256,512\}$. We find the performance of rating prediction is sensitive to $\lambda_r$, so we first search it from $\{10^{-1},...,10^{-4}\}$ and then from a finer range. For simulation data, the selected parameter sets are $\{D_e, D_{gru},\lambda_e,lr\}=\{32,32,0.0001,0.001\}$ for exposure model and for rating prediction $\{D_r,\lambda_r,lr,B\}=\{64,0.01,0.01,256\}$. For real-world data, the selected parameter sets are $\{D_e, D_{gru},\lambda_e,lr\}=\{32,32,0.0001,0.001\}$ for exposure model and $\{D_r,\lambda_r,lr,B\}=\{32,0.004,0.01,256\}$ for rating model. In our GMF rating prediction model, we choose identity mapping as activation function by performance on validation set.

In Movielens and Goodreads data, the increased variance problem for IPS estimator is very severe. So we also adjust propensity scores from our model into a reasonable range. We also use the technique of propensity clipping \cite{strehl2010learning} to clip propensity scores which is too high or too low for all kinds of de-biasing methods including Pop and PF. These procedures and clipping thresholds are decided by performance on validation set.

\section{Experiments}
Since the ideal test set for evaluation of de-biased rating prediction is a set where users rate randomly selected items and there is no real-world dataset that contains both a random test set and sequential information at our best knowledge, existing research mainly generate a skewed test set by adjusting user activity and item popularity \cite{liang2016causalrec,bonner2018causal}, but a simulated skewed test set still suffers user feedback-loop bias.

In our experiments, compared with previous work that focused on simulated data, we evaluate our model with all baselines both on simulated data and real-world data, $Movielens$ and $Goodreads$ in which a test set without user feedback-loop bias can be extracted. In the procedure of de-biasing rate prediction, exposure probability prediction and de-biased rate prediction share the same training and validation set, while one test set from biased observational data is used to evaluate the task exposure probability prediction and the other unbiased test set for the task de-biased rate prediction. 

\subsection{Baselines}
We compare our dynamic exposure model with two baseline models in previous work\cite{liang2016causalrec}: popularity model (Pop) and Poisson factorization model (PF). In the popularity model, the exposure probability is the portion of the users who have been exposed to the item. In the Poisson factorization model, the exposure probability follows Poisson distribution whose parameter is determined by latent embeddings of user and item\cite{gopalan2015scalable}. To evaluate the performance of de-biased learning, we use GMF as the model predictor and trained with evaluation functions without de-biasing (Naive) and with propensity scores estimated by Pop, PF and our model.

\subsection{Evaluation}

\subsubsection{Exposure Prediction}
For the task of exposure probability prediction, we evaluate its performance by the following matrix on the test set for exposure: Negative Log-Likelihood of the estimated exposure probability (NLL, smaller NLL means more accurate exposure probability). Recall@50 and NDCG@50 (bigger Recall@50 and NDCG@50 means more reasonable estimators) are ranking-based evaluation metrics of recommendation and their detailed definitions can be found in \cite{liang2016causalrec,liang2016modeling}.

\subsubsection{De-biased Rate Prediction}
For this task, we evaluate its performance by measuring the distance between the predicted rating matrix $\hat{Y}$ and the true rating matrix $Y$. In particular, we evaluate de-biased rate prediction with Mean Squared Error (MSE) and Mean Absolute Error (MAE) on an unbiased test set.

\begin{equation}
\nonumber
MSE(\hat{Y}, {Y}) = \frac{1}{NM} \sum_{i=1}^{N}\sum_{j=1}^{M} (Y_{i,j} - \hat{Y}_{i,j})^2,
\end{equation}

\begin{equation}
\nonumber
MAE(\hat{Y}, {Y}) = \frac{1}{NM} \sum_{i=1}^{N}\sum_{j=1}^{M} |Y_{i,j}-\hat{Y}_{i,j}|,
\end{equation}

where N is the number of users and M is the number of items, $Y_{i,j}$ means the rate on the i users for the j items.

\subsubsection{Recommendation Diversity}

Besides the evaluation of the accuracy of the predicted ratings, for the experiments on real-world data $Movielens$ and $Goodreads$, we want to measure the quality of the recommendation another way. To this effect, we evaluate the diversity of recommendations provided by the trained rating prediction model. With the trained model, we will generate a list of highest-rated items (The length of the list is 10 in Movielens and 20 in Goodreads). Then we calculate the Gini coefficient and average dissimilarity \cite{kunaver2017diversity}. The Gini coefficient measures the inequality among the frequency distribution of items recommended to all users, let $P_j,j=1,...,M$ be the popularity of each item with recommendation and sort them in non-decreasing order as $P_{(j)}$($P_{(j)} \leq P_{(j + 1)}$), the Gini coefficient is calculated as follows:

\begin{equation}
\nonumber
    G = \frac{\sum_{j=1}^{M} (2j - M - 1)P_{(j)}}{M\sum_{j=1}^{M}P_{(j)}}.
\end{equation}

\noindent Average dissimilarity measures the dissimilarity within the recommendation list of each user, which is defined as follows:
\begin{equation}
\nonumber
    D = \frac{\sum_{j=1}^{M}\sum_{j'=1}^{M}1-Similarity(c_j,c_{j'})}{2M(M-1)}.
\end{equation}

\noindent where $c_j$ is a vector represents the characteristic of item $j$ and $Similarity$ is a similarity measurement. We use the vector of genres in Movielens dataset as $c_j$ and use cosine similarity. Based on the definitions, when Gini coefficient is smaller and the average dissimilarity is bigger, the recommendation result is more diverse.

\subsection{Experiment on Simulation Dataset}

\subsubsection{Simulated dataset}
For simulation, we create a community that consists of 3000 users and 1000 items. The rating matrix is generated with the same forms of functions used in previous related paper \cite{chaney2018algorithmic}. And we tune the parameters to make the distribution of data closer to real data. The dimension of the latent variables ($\alpha_i$ and $\beta_j$) is 10. $\mu_{\alpha} \sim Dirichlet(20)$ and $\alpha_i \sim Dirichlet({\mu}_{\alpha})$ for each user. Similarly, $\mu_{\beta} \sim Dirichlet(100)$ and $\beta_j \sim Dirichlet({\mu}_{\beta})$ for each item. We first sample a value from Beta distribution with mean as $\alpha_i \beta_j^{T}$ and variance as $0.01$. Then we times the value with 10 to getting the rating $Y_{i,j}$ make. All ratings range in $[0,10]$. The true similarity matrix of items is defined as $S$ and $S_{j_1,j_2} = \beta_{j_1} \beta_{j_2}^{T}$. 

We use the following procedure to simulate the user feedback-loop: a fixed similarity matrix $\hat{S}$ is generated in advance to simulate the system have a pre-trained but not perfect recommendation algorithm: each $\hat{S}_{j_1,j_2}$ is sampled from a beta distribution with a mean of true similarity $S_{j_1,j_2}$ and variance of $0.01$. The system will combine a user's rating history and $\hat{S}$ to determine recommendation ranking with item-based collaborative filtering algorithms. The exposure distribution is affected by recommendation ranking: the exposure probability of top-100 ranked items will be 10 times as the others. For each user, the process will iterate for 30 times to generate the {\em biased observational data}. We choose the 1-20,21-25 and 26-30 items of each user as training, validation and test set for exposure probability prediction. We sample an extra 20 items randomly to construct the {\em unbiased test data} for rating prediction. The simulation will be repeated with ten random seeds. For each simulation "world", we will train each exposure model for once and report the averaged performance over the ten worlds. For de-biased rating prediction, we will repeat experiment 10 times for each world and report the result all averaged over the $10 \times 10$ experiments.

\subsubsection{Performance on Simulation Dataset}

The results of simulation experiments is summarized in Table ~\ref{table:expo_simu} and ~\ref{table:rate_simu}. As is illustrated in Table ~\ref{table:expo_simu}, for exposure prediction, our model outperforms the baseline methods Pop and PF obviously in the indicator of NLL, RECALL@50 and NDCG50. In Table ~\ref{table:rate_simu}, the result shows that Pop and PF have a slight performance improvement over Naive method on the simulation dataset, while our model surpasses all baselines for the task of de-biased estimator.

\begin{table}
  \caption{Exposure prediction performance on simualtion data.}
  \label{table:expo_simu}
  \centering
  \begin{tabular}{cccc}
    \hline
     & NLL & RECALL@50 & NDCG@50\\
    \hline
    Pop & $6.884 \pm 0.002$ & $0.0566 \pm 0.002$ & $0.023 \pm 0.001$ \\
    PF & $7.035 \pm 0.017$ & $0.0785 \pm 0.007$ & $0.032 \pm 0.003$ \\
    Ours & $\mathbf{6.813 \pm 0.019}$ & $\mathbf{0.1351 \pm 0.021}$ & $\mathbf{0.056\pm 0.008}$ \\
    \hline
  \end{tabular}
\end{table}

\begin{table}[htbp]
  \caption{Rating prediction performance on simualtion data.}
  \label{table:rate_simu}
  \centering
  \begin{tabular}{ccc}
    \hline
    & MSE & MAE\\
    \hline
    Naive & $2.001 \pm 0.066$ & $1.087 \pm 0.021$ \\
    Pop &  $1.990 \pm 0.035$ & $1.080 \pm 0.010$\\
    PF & $1.945 \pm 0.038$ & $1.065 \pm 0.010$\\
    Ours & $\mathbf{1.896 \pm 0.042}$ & $\mathbf{1.042 \pm 0.011}$ \\
    \hline
  \end{tabular}
\end{table}

\begin{table}
	\caption{Summary of each periods of Movielens data.}
	\label{table:summary_movielens}
	\centering
	\begin{tabular}{cccc}
		\hline
		Period & \# of users & \# of items & \# of records\\
		\hline 
		$P_1$ & 16627 & 1493 & 980617 \\
		$P_2$ & 16835 & 1669 & 982136 \\
		$P_3$ & 15717 & 1674 & 905040 \\
		\hline
	\end{tabular}
\end{table}

\begin{table}
	\caption{Summary of each periods of Goodreads data.}
	\label{table:summary_goodreads}
	\centering
	\begin{tabular}{cccc}
		\hline
		Period & \# of users & \# of items & \# of records\\
		\hline 
		$P_1$ & 10000 & 2970 & 642658 \\
		$P_2$ & 10000 & 2992 & 648177 \\
		$P_3$ & 10000 & 2999 & 651600 \\
		\hline
	\end{tabular}
\end{table}

\begin{table*}[htbp]
	\caption{Exposure prediction performance on Movielens data.}
	\label{table:obs_movielens}
	\centering
	\begin{tabular}{cccccccccc}
		\hline
		\multicolumn{1}{c}{} & \multicolumn{3}{c}{$P_1$} & \multicolumn{3}{c}{$P_2$}  & \multicolumn{3}{c}{$P_3$} \\
		\hline
		&  NLL & RECALL@50 & NDCG@50&  NLL & RECALL@50 & NDCG@50 &  NLL & RECALL@50 & NDCG@50\\
		\hline
		Pop & $6.563$ & $0.272$ & $0.151$ & $6.521$ & $0.285$ & $0.161$ & $6.489$ & $0.287$ & $0.163$ \\
		PF & $6.560$ & $0.282$ & $0.152$ & $6.563$ & $0.290$ & $0.158$ & $6.515$ & $0.297$ & $0.160$ \\
		Ours & $\mathbf{5.862}$ & $\mathbf{0.435}$ & $\mathbf{0.258}$ & $\mathbf{5.810}$ & $\mathbf{0.462}$ & $\mathbf{0.275}$ & $\mathbf{5.787}$ & $\mathbf{0.474}$ & $\mathbf{0.285}$ \\
		\hline
	\end{tabular}
\end{table*}

\begin{table*}[htbp]
	\caption{Rating prediction performance on Movielens data.}
	\label{table:rate_movielens}
	\centering
	\begin{tabular}{ccccccc}
		\hline
		\multicolumn{1}{c}{} & \multicolumn{2}{c}{$P_1$}  & \multicolumn{2}{c}{$P_2$} & \multicolumn{2}{c}{$P_3$} \\
		\hline
		&  MSE & MAE & MSE & MAE & MSE & MAE\\
		\hline 
		Naive  & $ 1.135 \pm 0.017$ & $0.803 \pm 0.005$ & $1.320 \pm 0.018$ & $0.863 \pm 0.005$ & $1.388 \pm 0.046$ & $0.886 \pm 0.012$\\
		Pop  & $1.149 \pm 0.017$ & $0.810 \pm 0.006$ & $1.311 \pm 0.017$ & $0.864 \pm 0.006$ & $1.388 \pm 0.062$ & $0.887 \pm 0.018$\\
		PF  & $1.125 \pm 0.015$ & $0.802 \pm 0.005$ & $1.300 \pm 0.020$ &$0.860 \pm  0.006$ & $1.373 \pm 0.047$ & $0.883 \pm 0.014$ \\
		Ours & $\mathbf{1.116 \pm 0.017}$ & $\mathbf{0.801 \pm 0.005}$ & $\mathbf{1.275 \pm 0.029}$ & $\mathbf{0.851 \pm 0.010}$ & $\mathbf{1.341 \pm 0.038}$ & $\mathbf{0.874 \pm 0.012}$ \\
		\hline
	\end{tabular}
\end{table*}

\begin{table*}[htbp]
	\caption{Diversity of recommendation results on Movielens data.}
	\label{table:diversity_movielens}
	\centering
	\begin{tabular}{ccccccc}
		\hline
		\multicolumn{1}{c}{} & \multicolumn{2}{c}{$P_1$}  & \multicolumn{2}{c}{$P_2$} & \multicolumn{2}{c}{$P_3$} \\
		\hline
		&  Gini & Avg-Dis & Gini & Avg-Dis & Gini & Avg-Dis\\
		\hline
		Naive  & $0.929 \pm 0.008$ & $0.633 \pm 0.007$ & $0.930 \pm 0.006$ & $0.639 \pm 0.009$ & $0.929  \pm 0.006$ & $0.631 \pm 0.010$\\
		Pop  & $0.913 \pm 0.010$ & $0.684 \pm 0.012$ & $0.908 \pm 0.009$ & $\mathbf{0.688 \pm 0.010}$ & $0.897 \pm 0.014$ & $\mathbf{0.690 \pm 0.013}$\\
		PF  & $0.904 \pm 0.014$ & $0.685 \pm 0.011$ & $0.899 \pm 0.014$ & $0.685 \pm 0.015$ & $0.895 \pm 0.012$ & $ 0.680 \pm 0.010$\\
		Ours & $\mathbf{0.883 \pm 0.014}$ & $\mathbf{0.692 \pm 0.006}$ & $\mathbf{0.890 \pm 0.012}$ & $0.686 \pm 0.013$ & $\mathbf{0.883 \pm 0.017}$ & $0.683 \pm 0.013$ \\
		\hline
	\end{tabular}
\end{table*}

\subsection{Experiments on Real-World Dataset}
In this section, we will introduce our experiments on $Movielens$ dataset and $Goodreads$ dataset. In each experiment, we will demonstrate the dataset pre-processing, the experiment settings, the results, and our analysis.

\subsubsection{Movielens Dataset}
\label{sec:movielens}
Movielens-20M is a benchmark dataset for movie rating prediction. The data ranging from 1996 to 2015 when the interface and recommendation algorithms used in Movielens website change \cite{harper2016movielens}. We study the range from 2005 to 2011 when Movielens system remains the same interface and recommendation algorithm. We further divide the data into three overlapped periods [01/01/2005,31/12/2008], [01/01/2006,31/12/2010] [01/01/2007,31/12/2011] (called $P_1$,$P_2$ and $P_3$ respectively) with similar amount of data to show our method can achieve consistent advantage over baselines. For each period, we select the users who registered in this period. And we only keep the users who rate at least 65 items and use the first 65 items for training and evaluation. 

We decide the split of the dataset based on the system's rule: a new user is asked to rate 15 items before the system recommends movies to him/her. So the first 15 items rated by a user will not be affected by user feedback-loop bias. So we select the first 15 items as a test set of rating prediction. We used the 16-45 and the 46-55 items for training and validation. The 56-65 items are used as the test set of exposure prediction. We only keep the items with popularity more than 100. To avoid leakage of information, We do not include the first 15 items when training the sequential exposure model. After pre-processing, the statistics of each period are summarized in Table ~\ref{table:summary_movielens}. For each period, we repeat the rate prediction experiments 10 times to ensure the results are reliable.

\subsubsection{Performance on Movielens Dataset}

\begin{table*}[htbp]
  \caption{Exposure prediction performance on Goodreads data.}
  \label{table:obs_goodreads}
  \centering
  \begin{tabular}{cccccccccc}
    \hline
    \multicolumn{1}{c}{} & \multicolumn{3}{c}{$P_1$} & \multicolumn{3}{c}{$P_2$}  & \multicolumn{3}{c}{$P_3$} \\
    \hline
     &  NLL & RECALL@50 & NDCG@50&  NLL & RECALL@50 & NDCG@50 &  NLL & RECALL@50 & NDCG@50\\
    \hline
    Pop & $7.589$ & $0.148$ & $0.085$ & $7.612$ & $0.141$ & $0.081$ & $7.510$ & $0.140$ & $0.803$ \\
    PF & $7.296$ & $0.168$ & $0.093$ & $7.279$ & $0.170$ & $0.093$ & $7.267$ & $0.164$ & $0.090$ \\
    Ours & $\mathbf{6.944}$ & $\mathbf{0.284}$ & $\mathbf{0.217}$ & $\mathbf{7.033}$ & $\mathbf{0.264}$ & $\mathbf{0.198}$ & $\mathbf{7.061}$ & $\mathbf{0.255}$ & $\mathbf{0.194}$ \\
    \hline
  \end{tabular}
\end{table*}

\begin{table*}[htbp]
	\caption{Rating prediction performance on Goodreads data.}
	\label{table:rate_goodreads}
	\centering
	\begin{tabular}{ccccccc}
		\hline
		\multicolumn{1}{c}{} & \multicolumn{2}{c}{$P_1$}  & \multicolumn{2}{c}{$P_2$} & \multicolumn{2}{c}{$P_3$} \\
		\hline
		&  MSE & MAE & MSE & MAE & MSE & MAE\\
		\hline 
		Naive  & $ 2.819 \pm 0.008$ & $1.345 \pm 0.006$ & $3.240 \pm 0.010$ & $1.470 \pm 0.005$ & $3.815 \pm 0.016$ & $1.611 \pm 0.005$\\
		Pop & $2.772 \pm 0.020$ & $1.340 \pm 0.006$  & $3.255 \pm 0.017$  & $1.476 \pm 0.004$  & $3.801 \pm 0.010$  & $1.619 \pm 0.012$ \\
		PF  & $2.770 \pm 0.011$  & $1.332 \pm 0.004$ & $3.243 \pm 0.017$ & $1.470 \pm 0.005$ & $3.822 \pm 0.039$ &  $1.622 \pm 0.018$ \\
		Ours & $\mathbf{2.766 \pm 0.028}$ & $\mathbf{1.328 \pm 0.010}$ & $\mathbf{3.227 \pm 0.012}$ & $\mathbf{1.465 \pm 0.008}$ & $\mathbf{3.798 \pm 0.012}$ & $\mathbf{1.609 \pm 0.007}$ \\
		\hline
	\end{tabular}
\end{table*}

\begin{table*}[htbp]
	\caption{Diversity of recommendation results on Goodreads data.}
	\label{table:diversity_goodreads}
	\centering
	\begin{tabular}{ccccccc}
		\hline
		\multicolumn{1}{c}{} & \multicolumn{2}{c}{$P_1$}  & \multicolumn{2}{c}{$P_2$} & \multicolumn{2}{c}{$P_3$} \\
		\hline
		&  Gini & Avg-Dis & Gini & Avg-Dis & Gini & Avg-Dis\\
		\hline
		Naive  & $0.958 \pm 0.017$ & $0.580 \pm 0.030$ & $0.954 \pm 0.010$ & $0.617 \pm 0.026$ & $0.940  \pm 0.010$ & $0.620 \pm 0.018$\\
		Pop  & $0.939 \pm 0.011$ & $0.586 \pm 0.025$ & $0.928 \pm 0.008$ & $0.646 \pm 0.015$ & $0.936 \pm 0.004$ & $0.642 \pm 0.008$\\
		PF  & $0.933 \pm 0.021$ & $0.562 \pm 0.046$ & $0.928 \pm 0.008$ & $0.632 \pm 0.025$ & $0.927 \pm 0.005$ & $ 0.642 \pm 0.009$\\
		Ours & $\mathbf{0.924 \pm 0.011}$ & $\mathbf{0.591 \pm 0.021}$ & $\mathbf{0.918 \pm 0.016}$ & $0.641 \pm 0.012$ & $\mathbf{0.926 \pm 0.006}$ & $\mathbf{0.649 \pm 0.015}$ \\
		\hline
	\end{tabular}
\end{table*}

\begin{figure}
	\centering
	\includegraphics[width=3in]{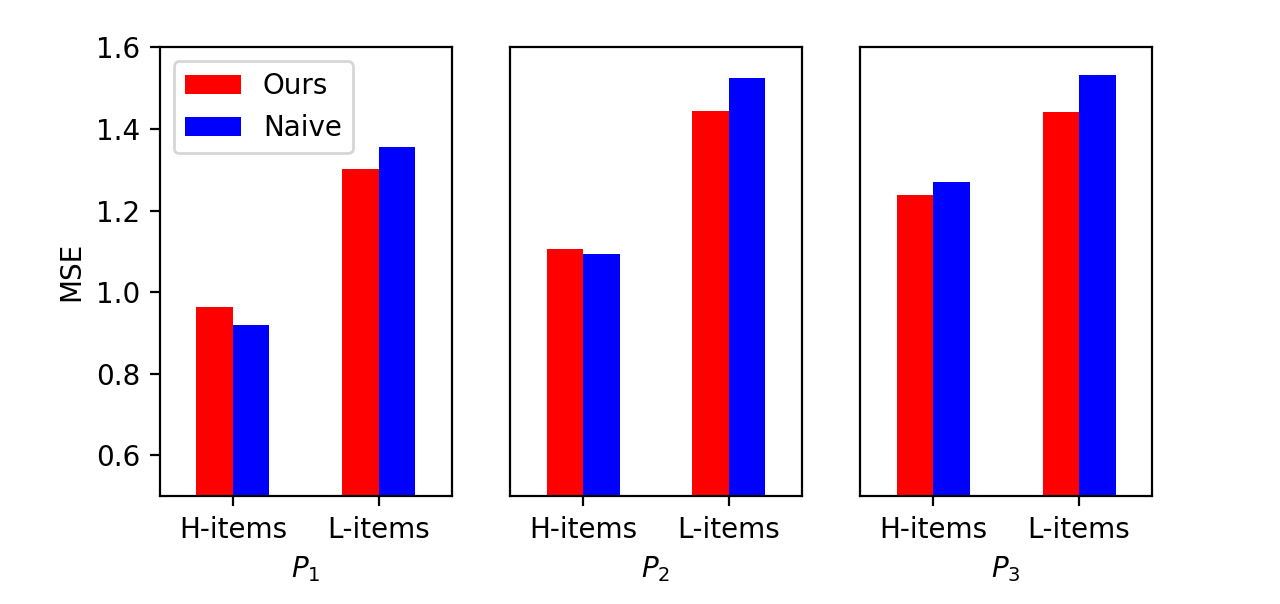}
	\caption{MSE of items with different popularities in Movielens data.}
	\label{fig:results}
\end{figure}

Table ~\ref{table:obs_movielens} shows the performance of exposure prediction in Movielens dataset. The performance of rating prediction and diversity of recommendation results are shown in Table ~\ref{table:rate_movielens} and ~\ref{table:diversity_movielens}. 
As we treat the selection bias sequentially, our model gets more accurate exposure probability and outperforms all baselines in the three periods in the tasks of exposure prediction. For the task of rating prediction, our model gets a better performance compared with the baselines. In the table ~\ref{table:diversity_movielens}, our model achieves more diverse recommendations. We also separate test data of rate prediction in half by items' popularity in training data, then compare MSE in high-popularity items (H-items) and low-popularity ones (L-items) in Figure~\ref{fig:results}. We can see the rating prediction on low-popularity items can benefit from our debiasing method.

\subsubsection{Goodreads Dataset}
Goodreads is a large-scale dataset from the book review website $GoodReads$ which contains 229,154,523 records and they are organized from 876,145 users' bookshelves and covers 2,360,650 books \cite{wan2018item}. In which we select the data collected after 15th September 2011 since then the $GoodReads$ recommendation system changed the recommendation algorithm. Similar to the Movielens recommendation system, the $GoodReads$ asks a new user who registered after 15th September 2011 to rate 20 books firstly before the system recommends customized books. We decide the first 20 books as test set of rating prediction on account of the first 20 items rated by a user will not be affected by user feedback-loop bias. We choose the 21-50 items for training, 51-60 for validation and 61-70 for test. To avoid information leakage, the training dataset doesn't contain the first 20 items when we train the sequential exposure model. For all records, we only select 3000 the most popular books.

As the records range from 2011 to 2017, we divide the data into 3 disjoint periods[16/09/2011,31/12/2012], [01/01/2013,31/12/2014] and [01/01/2015,31/12/2017](called $P_1$, $P_2$, $P_3$ respectively) with similar amount of records. For each period, we select the users who registered in this period. Considering too much time-consuming in each experiment if we use all records, we select 10000 records randomly each period in all the experiments. After pre-processing, the statistics of each period are summarized in Table ~\ref{table:summary_goodreads}. For each period, we repeat the rate prediction experiments 10 times as the Movielens experiment does.

\subsubsection{Performance on Goodreads Dataset}
The results of the experiments on Goodreads Dataset are shown in Table ~\ref{table:obs_goodreads}, ~\ref{table:rate_goodreads} and ~\ref{table:diversity_goodreads}. In which the exposure prediction performance is in Table ~\ref{table:obs_goodreads}, our model surpasses all baseline in each period. As for the accuracy of rating predictions, Table ~\ref{table:rate_goodreads} illustrates that all the methods perform almost the same. As there may exist bias-variance trade-off in de-biasing methods \cite{schnabel2016recommendations}, more accurate exposure probability predictions may mean higher variance of propensity scores. The high variance of propensity is much severe in real-world data and it can hurt the performance of the debiased method.

Since the category of one book is indistinct, the book may belong to the category of romance, fantasy and story. In our task of measuring dissimilarity, we regard the most relevant category of the book as its genre and the results are shown in Table ~\ref{table:diversity_goodreads}. We can see that our method provides the most diverse recommendation results.

\subsection{Discussion}
Based on the experiments on simulated data and real-world data, we make the following observations:

\begin{enumerate}
\item Our sequential exposure model substantially and significantly outperforms the static models. In the real-world datasets, our model achieves the best performance on the task of exposure probability prediction. In the task of rating prediction, we surpass existing de-biasing methods in simulated dataset. This suggests there exists user feedback-loop bias in real-world data and models with the dynamic mechanism can improve the performance of exposure prediction.  
\item In simulation and Movielens data, the rating prediction with de-biasing mechanism is better than the naive estimator, while our method achieves the best results. We can also see recommendation performance is more stable across popularities of items from Figure~\ref{fig:results}. In Goodreads dataset, the performance between naive estimator and debiased estimators are quite similar due to the negative impact of high variance counterbalances the postive impact of less biased estimation. 
\item Our dynamic model improves the recommendation quality. With de-biasing learning, the recommendation results are more diverse with the measurement of both Gini coefficient and average dissimilarity. In the Movielens and Goodreads experiments, our method outperforms baselines by Gini coefficient in all periods and achieves close average dissimilarity compared to baselines. By propensity score re-weighting, more diverse items are recommended when we get better results on rating prediction.  
\end{enumerate}

\section{Conclusions}
We propose to systematically correct user-feedback loop bias for training and evaluating explicit-rating recommendation algorithms. We design a novel probabilistic graphical model structure to estimate the dynamic item exposure probability distributions with respect to individual users. The estimated probabilities are used as propensity scores to adjust the user feedback-loop bias in sequential rating prediction. We show the superiority of our method over other models with statically estimated exposure probabilities. Experiments on simulated data demonstrate the existence of user feedback bias and our method can achieve better performance on exposure probability prediction and de-biasing rating prediction. By adjusting the bias sequentially, more diverse items are recommended with more reasonable exposure probability. More importantly, we get expected results on exposure predictions and de-biasing rating predictions on the two real-world datasets.

\begin{acks}
\end{acks} 

\bibliographystyle{ACM-Reference-Format}
\bibliography{sample-base}

\appendix

\end{document}